\documentclass[conference, 10pt]{IEEEtran}
\IEEEoverridecommandlockouts

\usepackage[utf8]{inputenc} 
\usepackage[T1]{fontenc}
\usepackage{url}              
\usepackage{cite}             

\usepackage[cmex10]{amsmath}  
\usepackage{mleftright}       
\mleftright                   
\usepackage{amsthm}
\usepackage{graphicx}         
\usepackage{booktabs}         

\usepackage{cite}
\usepackage{amsmath}
\usepackage{amssymb}
\usepackage{algorithmic}
\usepackage{textcomp}
\usepackage{xcolor}
\def\BibTeX{{\rm B\kern-.05em{\sc i\kern-.025em b}\kern-.08em
    T\kern-.1667em\lower.7ex\hbox{E}\kern-.125emX}}
\usepackage{graphicx}
\usepackage{subfiles}
\usepackage[inline]{enumitem}

\usepackage{array} 
\usepackage{url}
\usepackage{epstopdf}
\usepackage{tikz}
\usetikzlibrary{automata,arrows,positioning,calc,
shapes,chains}
\usepackage{mathtools}
\usepackage{texdef2015}
\allowdisplaybreaks

\newcommand{\Lcal}{\mathcal{L}}
\newcommand{\Qcal}{\mathcal{Q}}

\newcommand{\R}{\mathbb{R}}

\newcommand{\ql}{q_l}

\newcommand{\laml}[1][l]{\lambda^{(#1)}}

\newcommand{\rowvec}[1]{[\begin{matrix} #1\end{matrix}]}
\newcommand{\rvec}[1]{\smrvec{#1}}
\newcommand{\smrvec}[1]{\setlength\arraycolsep{2pt}\rowvec{#1}}

\newcommand{\piv}{\text{\boldmath{$\pi$}}}

\newcommand{\pibar}{\bar{\pi}}
\newcommand{\vvbar}{\bar{\vv}}

\newcommand{\onev}[1][n]{{\mathbf{1}}_{#1}}
\renewcommand{\vec}[1]{\begin{bmatrix} #1\end{bmatrix}}

\newlength{\swwidth}

\newcommand{\lam}{\lambda}

\newcommand{\sig}{\sigma}

\newcommand{\oneage}{x_1}
\newcommand{\twoage}{x_2}
\newcommand{\maxage}{\hat{x}}
\newcommand{\dpage}{\hat{y}}
\newcommand{\monage}{\hat{z}}

\begin{document}

\title{Timely Processing Of Updates From Multiple Sources} 


\author{\IEEEauthorblockN{Vishakha Ramani,
Ivan Seskar,
Roy D. Yates}
\IEEEauthorblockA{WINLAB, Rutgers University \\
Email: \{vishakha, seskar, ryates\}@winlab.rutgers.edu
}}

\maketitle

\begin{abstract}
We consider a system where the updates from independent sources are disseminated via a 
publish-subscribe mechanism. The sources are the publishers
and a decision process (DP), acting as a subscriber, derives decision updates from the source data.
We derive the stationary expected age of information (AoI) of decision updates delivered to a monitor. 
We show that a lazy computation policy in which the DP may sit idle before computing its next decision update can reduce the average AoI at the monitor even though the DP exerts no control over the generation of source updates. 
This AoI reduction is shown to occur because lazy computation can offset the negative effect of high variance in the computation time.  
\end{abstract}

\section{Introduction}
A dense metropolis is a complex traffic environment, and autonomous cars with a plethora of attached sensors still have limited situational awareness. Hence, holistic situational awareness of a cloud-connected vehicle is facilitated by
\begin{enumerate}
    \item timely collection of  sensory inputs from different sources (e.g., other vehicles, pedestrians, and smart city infrastructure sensors)
    , and
    \item employing this sensor data 
    to provide timely feedback or decision updates to participating mobile clients. 
\end{enumerate}
For example, cameras at a smart-city intersection \cite{Cosmos} can capture video or images of the intersection and send this data to a processing system at the edge. The system can then analyze the data to identify and inform potential public safety risks such as vehicles running red lights or pedestrians crossing the street outside of designated areas.

Although, it is possible that the system delivers source updates to interested clients using traditional synchronous request/reply communication paradigm, the rigid structure of such interaction renders the system inefficient for real-time decisions \cite{Carzaniga-pubsub}.
Publish-Subscribe (Pub-Sub) systems are an 
alternative communication paradigm 
that enables efficient and scalable communication among various components. The delivery of information from publishers to subscribers is decoupled, meaning that they need not  be aware of each other's existence.
For example, cameras can publish video frames to the system while a processing system subscribes to this data and uses it to make real-time decisions about intersection safety, such as adjusting the timing of traffic lights or alerting drivers to pedestrians in the area.
A middleware (also known as a broker) acts as an intermediary 
to manage this distribution of information from publishers to subscribers.

The presence of different types of subscribers with varying time scales of operation in an edge computing system allows for flexibility and adaptability to different requirements and use cases. 
For instance, a real-time subscriber could be an analytics module that processes sensor data in real-time to detect anomalies or trigger immediate actions. On the other hand, a batch processing subscriber could be a processing module that handles a batch of sensor data collected over a certain time window to generate periodic reports or perform long-term trend analysis.
Consequently, such heterogeneity requires decoupling subscribers from publishers. 

In order to facilitate the required information dissemination,
the data is pushed to the middleware from publisher. However, 
two complementary communication modes between middleware and subscriber exist: push and pull modes. Under the push mode, the subscriber passively receives information from the pub-sub middleware. In a pull-based approach, subscribers request messages from the middleware when they are ready to receive them, rather than waiting for the middleware to push messages to them.

There are pros and cons to both approaches. However, the pull mode is better suited for handling a diverse range of subscribers, without requiring a broker to determine the data transfer rate for each of them.
Subscribers have more control over the rate at which they consume messages, making it easier to manage their individual needs \cite{kafka}. 


Along with this, a real-time database system, acting as a middleware system,
can support information dissemination between publishers
and subscribers.
With these design choices in mind, we present an analytical framework 
based on the publish-subscribe interaction paradigm, that is to be used for disseminating source updates to interested clients that process these source updates and arrive at decisions in a timely way.

\begin{figure}[t]
    \centering
    \includegraphics[width = \linewidth]{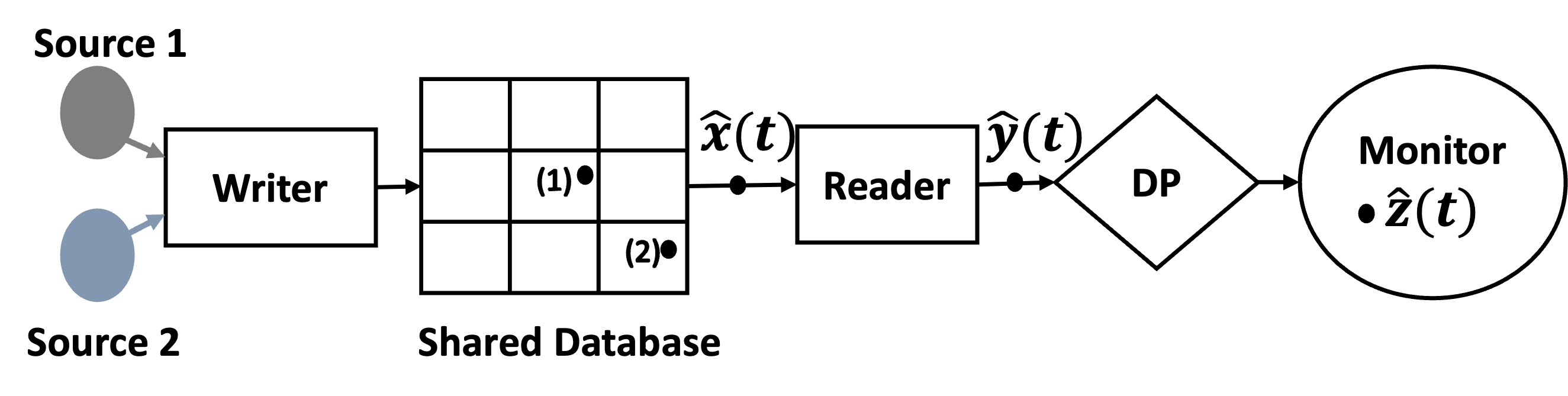}    \caption{A writer updates shared database with information fetched from two external sources. A decision process (DP) requests a reader process to read the pair of source updates from the memory. Monitors that track the age of source $1$ and $2$ updates  in the memory are denoted $\bullet(1)$ and $\bullet(2)$ respectively; $\bullet(\maxage(t))$ tracks the age of max-age process in the memory, $\bullet(\dpage(t))$ tracks the age of sampled max-age process, and $\bullet(\monage(t))$ tracks the age of computed decision updates at the external monitor.}
    \label{fig:tc-sys-model}
\end{figure}

\textbf{Considered Problem: } 
The freshness of status information received by subscriber plays an important role in decision making. 
In this work, we focus on a  fundamental problem: What is the average Age of Information (AoI) \cite{Kaul-YG-infocom2012} of  decision updates that are computed from  time-varying set of sensor data published in the database\footnote{Going forward, we refer to the database as {\em shared memory}, or simply as {\em memory}.}?
We model a  class of systems (see Fig.~\ref{fig:tc-sys-model}) in which two independent sources submit time-stamped updates to a  writer that is responsible for publishing the source measurements as updates in the memory.
A decision process (DP), as a subscriber, reads the pair of source $1$ and source $2$ updates from memory and derives a computational result, a {\em decision update}, from this pair that is delivered to a monitor. 

\textbf{Related Work: }
The issue of update timeliness using Age of Information (AoI) metric has been extensively studied; see the surveys  \cite{kosta2017age,Yates-SBKMU-2021jsac-survey} and references therein. A majority of studies focus on analyzing the average AoI of different sources in single-server multi-source queueing models under different service policies \cite{moltafet-multi-source, Yates-Kaul-isit2012, Huang-Modiano-isit2015}. 

More recently, there have been efforts to employ age optimization in diverse applications such as edge cloud processing for low-latency edge-assisted applications \cite{Zhong-ZYGZ-aoi2019, Yates-THR-infocom2017}, and timely mobile routing \cite{ramani-cyINFOCOM, ramani-cy-AoI-INFOCOM}. 
Authors in \cite{Zhong-ZYGZ-aoi2019} discuss a greedy traffic scheduling policy that selects the next processing request (job) that offers maximum age penalty reduction. \cite{Yates-THR-infocom2017} provides an analytical framework for the problem of optimizing frame rate and lag synchronization of server and player in a real-time cloud-assisted gaming application. 
Authors in \cite{ramani-cyINFOCOM, ramani-cy-AoI-INFOCOM} study the effects of concurrency constructs on timely updating of shared data structures used in network software and how this, in turn, affects timely routing of information updates.

In previous works, the system models assumed a tight-coupling between the source and monitor. The primary goal was to maintain the status information of the physical process at the destination nodes. For instance, the just-in-time update policy \cite{Kaul-YG-infocom2012} generated a fresh update instantaneously by the source, which started serving as soon as the current update in service was delivered to the destination node. Many works \cite{Zhou-Saad-arxiv2018, Bedewy-SS-arxiv2017old, Yates-isit2015, Sun-UBYKS-IT2017UpdateorWait} have extended upon this idea and aimed to find optimal policies for determining sampling times and updating processes to minimize the Age of Information (AoI) at the destination node.

In publish-subscribe systems with decoupled producers and consumers, where producers use independent status updating policies, and consumers compute on this status information, a key open problem is to study optimal policies for minimizing the Age of Information (AoI) of computed updates. This work aims to study how the decoupling of publishers and subscribers affect the
timeliness of updates computed 
from base set of sensor (source) updates.


\section{System Overview}
\label{sec:sysmodel}

In this work, we employ a model that captures  the asynchronous operation of the writer and reader of updates in the shared memory Pub-Sub system. 
There are three aspects to the system depicted in Fig.~\ref{fig:tc-sys-model}: 
\begin{enumerate*}
    \item writing the time-varying data received from two sources into the memory,
    \item the arbitration between reader and writer to access memory,
    \item reading the source data from memory and generating a decision update.  
\end{enumerate*}
We now give a brief overview of the writing, reading and decision computation processes.

\subsection{Writing source updates to the memory}
We assume each source $i \in \set{1,2}$ independently submits updates as a rate $\lam_i$ Poisson process to the network and 
that these updates arrive fresh at the writer, i.e. with age $0$. 
The write operations to memory have independent exponential $(\mu)$ service times.
We model the writer as a buffer-less service facility with blocking discipline.  Under this model, a source update arriving at the writer will be served only if the writer is idle; otherwise, the update is discarded.

\textit{Remark 1: } In the present study, we investigate a computational regime characterized by relatively longer decision update times compared to the write times of any update in the memory. Our focus is not on regimes where writing to the memory is the overloaded process. Instead, we are primarily interested in examining the delays associated with computational processing. Whether we adopt a buffer-less or a queuing model, the impact of queuing at the writer is expected to be minimal.


\subsection{Read-Copy-Update}
We assume that arbitration between the reader and writer for the shared memory is facilitated by Read-Copy-Update (RCU) mechanism. RCU is a lock-less concurrency construct that allows concurrent forward progress for both reader and writer. \cite{MckenneyRCU2001}. RCU can be broadly described in two steps \cite{kernelRCU}: 
\begin{enumerate*}
     \item To publish a newer version of a data item,
     the writer creates a copy of the RCU protected data item, modifies this copy with the newer version of this data item, and atomically replaces the old reference with a reference to this newer version.
     This publishing process runs concurrently 
     with ongoing read processes that continue to read the old copy/version using the old reference. However, new read requests 
     read the most recent 
     version. 
    \item Since some readers in progress hold reference to ``stale'' data, the system defers memory reclamation of old data until after each reader in progress has finished executing its read-side critical section. 
\end{enumerate*}

\textit{Remark 2: } RCU read operations can be performed concurrently without any locks, allowing for high concurrency and low contention. This can be particularly useful in systems with multiple subscribers and few publishers. In combination, RCU and Pub-Sub can enable efficient and scalable communication 
where multiple components need to access and update shared data in a concurrent and asynchronous manner.  


\subsection{Computing decision updates}\label{DPoverview}
%
We view the decision process (DP) reader as one of many subscribers to the updates in the memory system. 
The DP reader becomes aware of fresher updates in the memory only when it chooses to query the memory for a fresh sample of the source update pair. 
We assume a reader can fetch the updates of both source $1$ and source $2$ from memory in negligible time\footnote{This assumption is consistent with RCU reads being lightweight and fast, so that the heavier load is indeed induced by actual decision computation. Further, our model assumes that the DP reader 
fetches
updates from the memory 
at some finite average rate such that the combined read request process of all subscribers does not overload the shared memory system.}. 
With this assumption, the DP reader  is an observer that is {\em sampling} the pair of source updates from the memory as a point process. 
Based on this sample, the decision process derives a decision update  which is sent to the monitor, as shown  in Fig.~\ref{fig:tc-sys-model}. The reader process fetches the next sample of update pair from the memory only after the computation in progress is completed. 

\begin{figure}[t]
\centering
\begin{tikzpicture}[scale=0.18] 
\draw [<-|] (0,18) node [above] {age} -- (0,0) -- (16,0);
\draw [|->] (17,0) -- (38,0) node [right] {$t$};
\draw [very thick, dotted] (0,2) -- (4,6) -- (4,4) 
-- (9,9)  -- (9,1) -- (13,5) -- (13,3) -- (15,5);
\draw [very thick, dotted] (18,5) -- (22,9) -- (22,2) -- (27,7) -- (27,4) -- (31,8) -- (31,3) -- (35,7) -- (35,4) -- (37,6) node [right] {$\maxage(t)$};
\draw [very thick, dashed] (0,3) -- (6,9) -- (6,6) -- (14,14) -- (14,4);
\draw [very thick, dashed] (19,9) -- (19,6) -- (23,10) -- (23,3) -- (32,12) -- (32,4) -- (37,9) node [right] {$\hat{y}(t)$};
\draw [very thick] (0,6) -- (3,9) -- (3,6) -- (10,13) -- (10,10) -- (14,14);
\draw [very thick] (18,8) -- (20,10) -- (20,7) -- (29,16) -- (29,9) -- (37,17) node [right] {$\monage(t)$};
\draw 
(0,1) node [left] {$\hat{x}(0)$}
(0,3) node [left] {$\hat{y}(0)$}
(0,5) node [left] {$\hat{z}(0)$}
(6,0) node {$\scriptstyle\blacktriangledown$}
(6,0) node [below] {$\tau_1$} 
(14,0) node {$\scriptstyle\blacktriangledown$}
(14,0) node [below] {$\tau_2$}
(19,0) node {$\scriptstyle\blacktriangledown$}
(19,0) node [below] {$\tau_{i-2}$}
(23,0) node {$\scriptstyle\blacktriangledown$}
(23,0) node [below] {$\tau_{i-1}$}
(32,0) node {$\scriptstyle\blacktriangledown$}
(32,0) node [below] {$\tau_{i}$};
\draw  [|<->|] (0,-2.75) to node [below] {$Y_1$} (6,-2.75);
\draw  [|<->|] (6,-2.75) to node [below] {$Y_2$} (14,-2.75);
\draw  [|<->|] (6,-6) to node [below] {$T_2$} (10,-6);
\draw  [|<->|] (10,-6) to node [below] {$W_2$} (14,-6);
\draw  [|<->|] (19,-2.75) to node [below] {$Y_{i-1}$} (23,-2.75);
\draw  [|<->|] (23,-2.75) to node [below] {$Y_i$} (32,-2.75);
\draw  [|<->|] (23,-6) to node [below] {$T_i$} (29,-6);
\draw  [|<->|] (29,-6) to node [below] {$W_i$} (32,-6);
\draw [|<->|,blue] (25.8, 3) to node [left] {\small $H_i\!$} (25.8, 10);
\end{tikzpicture} \\
\caption{\small Example AoI evolution of the max-age process $\maxage(t)$ at the memory, the sampled max-age process $\hat{y}(t)$ with {\em lazy sampling} at the input to the DP, and the age process $\monage(t)$ at the monitor. The DP reader samples updates from the memory
at times $\tau_1, \tau_2, \ldots$, marked by $\scriptstyle\blacktriangledown$. $Y_i$ is the sampling period for sample $i$, $T_i$ is the computation time for decision update based on sample $i-1$, and $W_i$ is the waiting time to get the $i^{th}$ sample.}
\label{fig:tc-age-ev}
\end{figure}
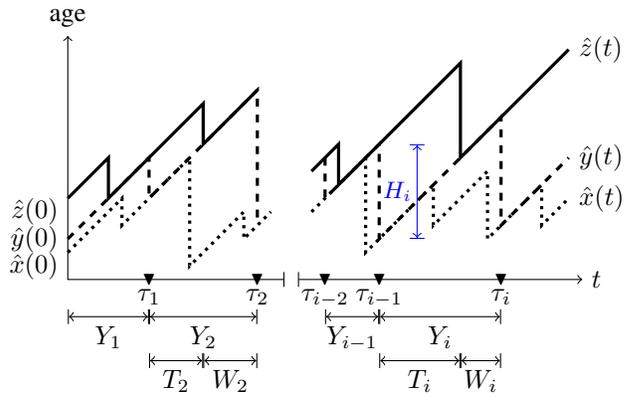

When the DP reader's inter-sample times form a renewal process, this is an example of the model of renewal process sampling of updates introduced in \cite{Yates-IT2020}.  In this model, the DP reader generates  an age process $\dpage(t)$  at the input to the DP that is a sampled version of the max-age process $\maxage(t)$ in the memory. Specifically, in the absence of a read, $\dpage(t)$ continues to grow at unit rate. However, if the DP reader makes a read at time $\tau$, then $\dpage(t)$ is reset to $\dpage(\tau)=\maxage(\tau)$. This update pair is then processed by the DP  for a time $T$ so that at time $\tau+T$ a decision update with age $\dpage(\tau)+T$ is delivered to the monitor. The age at the monitor, $\monage(t)$,  is  then reduced to $\monage(\tau+T)=\maxage(\tau)+T$. At this time, the DP reader may choose to fetch a new sample pair from the memory, or it may choose to wait for a time $W$ before fetching the next sample pair. When the DP reader employs non-zero waiting times, we say the DP is using a {\em lazy sampling} policy \cite{Yates-isit2015}. 
Fig.~\ref{fig:tc-age-ev} illustrates the evolution of age processes $\maxage(t), \dpage(t)$, and $\monage(t)$.

\subsection{Paper Overview and Contributions} 
We 
divide our AoI analysis into two stages: 1) We analyze the average age of updates in shared memory.  2) We analyze the additional delay induced by the  decision process computations.  
First, section \ref{sec:SHS} presents a stochastic hybrid system (SHS) evaluation of  the update age processes in the memory. For the system with sources $i=1,2$, we derive the stationary expected ages $\E{x_i(t)}$ as well as the expected age of the {\em max-age} process $\maxage(t)=\max(x_1(t),x_2(t))$, . 

In section~\ref{sec:decision-processing}, stage two of our analysis, we 
evaluate the age $\monage(t)$ of the decision update process at the monitor. 
The decision process is said to be {\em sampling} the source updates from the memory as it holds a sample of updates that were written to the memory.
Even though the sampling and computation of the DP makes no attempt to use the age of its sampled updates to optimize its operation, 
we show that a lazy sampling policy  will be able to reduce  $\monage(t)$. Here we will see that analysis of $\monage(t)$ is separable from the prior SHS analysis of the max-age process $\maxage(t)$  in the shared memory. In particular, the AoI reduction afforded by lazy sampling can be applied to any stationary update age process that is sampled by the DP. 

\section{Age of source updates in the memory} \label{sec:SHS}
Let $U_{i,1}, U_{i, 2} \ldots$ be the sequence of source $i$ update publication times. At any time $t$, $N_i(t)$ source $i$ updates have been published in the memory, and the most recent update is published at time $U_{i, N_i(t)}$.  It follows that the source $i$ update process has age
 $   x_i(t) = t - U_{i, N_i(t)}$
in the memory.
Under this model, the update age $x_i(t)$ is reset to the write time $W \sim \exp(\mu)$ when it is published at time $U_{i,N_i(t)}$. 
When the writer writes a fresh source $i$ update at time $t'$, the max-age process $\maxage(t)$ is reset to $\hat{x}(t') = x_j(t')$, with $j \neq i$. In the following, we use a Stochastic Hybrid System (SHS) to capture the evolution of update age processes in the memory.

\subsection{SHS Overview}
To evaluate AoI of source updates, we use a Stochastic Hybrid Systems (SHS)~\cite{hespanha2006modelling} approach,  a technique introduced for AoI evaluation in \cite{yates2018ToIT} and since employed in AoI evaluation of a variety of status updating systems \cite{Yates-aoi2018,Farazi-KB-aoi2018,Maatouk-AE-aoi2019,Kaul-Yates-isit2018priority,Maatouk-AE-ToN2020,Yates-IT2020,Moltafet-LC-CommLetters2021,Moltafet-LC-ISWCS2021}.
A stochastic hybrid system has a state-space with two components -- a  discrete component $q(t) \in \Qcal = \{0, 2, \ldots, M\}$ that is a continuous-time finite-state Markov Chain and a continuous component $\xv(t) = [x_0(t), \ldots , x_n(t)] \in \R^{n+1}$. In AoI analyses using SHS, each $x_{j}(t) \in \xv(t)$ describes an age process of interest. Each transition $l \in \Lcal$ is a directed edge $(q_l, q'_l)$ with a transition rate $\laml$ in the Markov chain. The age process vector evolves at a unit rate in each discrete state $q \in \Qcal$, i.e., $\frac{d\xv}{dt} = \dot{\xv}(t) = \onev$. A transition $l$ causes a system to jump from discrete state $q_l$ to $q_l^\prime$ and resets the continuous state from $\xv$ to $\xv'$ using a linear transition reset map $\Amat_l \in \{0,1\}^{(n \times n)}$ such that $\xv' = \xv \Amat_l$. For simple queues, examples of transition reset mappings $\set{\Amat_l}$  can be found in \cite{yates2018ToIT}.

For a discrete state $\qbar \in \Qcal$, let 
\begin{align}
    \Lcal_{\qbar} &= \{l \in \Lcal : q'_l = \qbar \}, &
    \Lcal'_{\qbar} &= \{l \in \Lcal : q_l = \qbar \}.
\end{align}
denote the respective sets of incoming and outgoing transitions.
Age analysis using SHS is based on the  expected value processes $\set{\vv_q(t)\colon q\in \Qcal}$ such that 
 $\vv_q(t)
 =\E{\xv(t)\delta_{q,q(t)}}$,
 with $\delta_{i,j}$ denoting the Kronecker delta function. For the SHS models of age processes considered here, each $\vv_q(t)$ will converge to a fixed point $\vvbar_q$. The fixed points $\set{\vvbar_q\colon q\in\Qcal}$ are the solution to a set of age balance equations. The following theorem provides a simple way to calculate the age-balance fixed point and then the average age. 
\begin{theorem}\thmlabel{AOI-SHS}
\cite[Theorem~4]{yates2018ToIT}
If the discrete-state Markov chain $q(t)\in\Qcal=\set{0,\ldots,M}$ is ergodic with stationary distribution 
$\bar{\piv}=\rvec{\bar{\pi}_0 &\cdots &\bar{\pi}_M}>0$ and there exists a 
non-negative vector $\vvbar=\rvec{\vvbar_0&\cdots&\vvbar_M}$
such that 
\begin{align}
\bar{\vv}_{\qbar}\sum_{l\in\Lcal_{\qbar}}\laml &=\onev[]\bar{\pi}_{\qbar}+ \sum_{l\in\Lcal'_{\qbar}}\laml \bar{\vv}_{\ql}\Amat_l,\quad \qbar\in\Qcal,\eqnlabel{AOI-SHS-v}
\end{align}
then
the average age vector is 
$\E{\xv}\!=\!{\displaystyle \limty{t}}\E{\xv(t)}\!=\!
\sum_{\qbar\in\Qcal} \vvbar_{\qbar}$.
\end{theorem}

\subsection{SHS Analysis of Age in Shared Memory}
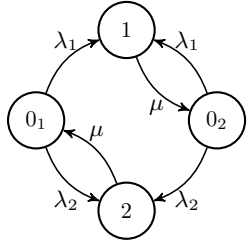
\begin{figure}[t]
$\setlength{\extrarowheight}{0.5mm}
\begin{array}[b]{cccc}
l & q_l\to q^\prime_l & \laml &  \xv\Amat_l \\ \hline
1 & 0_1\to1 & \lam_1 & \rvec{0, x_1, x_2, \hat{x}} \\
2 & 1\to0_2 & \mu & \rvec{x_0, x_0, x_2, x_2} \\
3 & 0_2\to1 & \lam_1 & \rvec{0, x_1, x_2, \hat{x}} \\
4 & 0_2\to2 & \lam_2 & \rvec{0, x_1, x_2, \hat{x}} \\
5 & 2\to0_1 & \mu & \rvec{x_0, x_1, x_0, x_1} \\
6 & 0_1\to2 & \lam_2 & \rvec{0, x_1, x_2, \hat{x}}
\end{array}$
\hfill
\begin{tikzpicture}[->, >=stealth', auto, semithick, node distance=2cm]
\tikzstyle{every state}=[fill=none,draw=black,thick,text=black,scale=0.85]
\node[state]    (0)                     {$0_1$};
\node[state] (1)[above right of=0] {$1$};
\node[state] (2)[below right of=0] {$2$};
\node[state] (3)[below right of=1] {$0_2$};
\path
(0) edge[bend left=25, above] node{\small $\lam_1$}  (1)
(0) edge[bend right=25, below] node{\small $\lam_2$} (2)
(1) edge[bend right=25, below] node{\small $\mu$} (3)
(3) edge[bend right=25, above] node{\small $\lam_1$} (1)
(3) edge[bend left=25, below] node{\small $\lam_2$} (2)
(2) edge[bend right = 25,above] node{\small $\mu$} (0);
\end{tikzpicture}
\caption{The SHS transition/reset maps and Markov chain for the update age in the shared memory.}
\label{fig:simplified-tc-mc}
\end{figure}
The age of updates in a shared memory system 
with bufferless service at the writer  can be described by the SHS Markov chain and table of state transitions shown in Fig.~\ref{fig:simplified-tc-mc}.  The continuous age state vector 
is $\xv = \rvec{x_0, x_1, x_2, \hat{x}}$, where 
$x_0$ is the age of the update being written; $x_i$, $i=1,2$,  is the age of the source $i$ update in memory; and
$\hat{x}=\max(x_1,x_2)$. 
The discrete state is $\Qcal = \set{0_1, 0_2, 1, 2}$. At time $t$, the system  is in state $0_i$ if the writer is idle and the oldest update belongs to source $i$. State $i\in \set{1, 2}$ corresponds to the writer writing source $i$ update.

We now describe SHS transitions enumerated in the table in Fig.~\ref{fig:simplified-tc-mc}. 
For each collection of transitions, we focus on the age state components 
that change.
\begin{itemize}
\item {$l=1,3,4,6$:} In system idle states $0_1$ and $0_2$, the writer receives a new source update and initiates a new write mechanism. $x_0^\prime = 0$ as the writer receives a fresh update, and $x_1^\prime , x_2^\prime, \hat{x}^\prime$ are unchanged as the update is not yet written to the memory. 

\item {$l=2, 5$:} The writer finishes writing and publishes a new source update. 
    \begin{description}
        \item[$l=2$]: the writer publishes source $1$ update: $x_1^\prime = x_0$ as the age of source $1$ update in the memory is reset to just written update. The source $2$ update becomes the oldest update in the memory; hence, 
        $\hat{x}^\prime = x_2$. 
        \item[$l=5$]: The writer publishes source $2$ update: $x_2^\prime = x_0$, the source $1$ update becomes the oldest update, and $\hat{x}^\prime = x_1$.
    \end{description} 
\end{itemize}
\noindent For the SHS analysis, we  employ  the normalized rates \begin{equation}
    \rho_1={\lam_1}/\mu,\quad \rho_2={\lam_2}/{\mu}.
\end{equation} 
We note that $\rho = \rho_1 + \rho_2$ is the total offered load of source updates being written to the memory. 
The Markov chain in Fig.~\ref{fig:simplified-tc-mc} has stationary probabilities $\piv$
with normalization constant $C_\pi$ given by
\begin{subequations}
\begin{align}
    \piv &=  \rvec{\pi_{0_1} & \pi_1 & \pi_2 & \pi_{0_2}}
    = C_\pi^{-1} \rvec{\rho_2 / \rho & \rho_1 & \rho_2 & \rho_1/ \rho},\\
C_\pi &= 1+\rho.
\end{align}
\end{subequations}
With the shorthand notation 
\begin{equation}
    \lam=\lam_1+\lam_2,
    \end{equation}
we now use \Thmref{AOI-SHS} to solve for
\begin{align}
    \vvbar &= \rvec{\vvbar_{0_1}&\vvbar_{1}&\vvbar_{2}&\vvbar_{0_2}},
\end{align}
where 
$\vv_q=\rvec{v_{q0}&v_{q1}&v_{q2}&v_{q3}}, \forall q \in \Qcal$. 
This yields
\begin{subequations}
\eqnlabel{simplified-tc-shs-eqn}
\begin{align}
\lam \vvbar_{0_1} &= \onev[]\pibar_{0_1} + \mu \vvbar_2 \Amat_5, \\
\mu\vvbar_1 &= \onev[]\pibar_1 + \lam_1 \vvbar_{0_1} \Amat_1 + \lam_1 \vvbar_{0_2} \Amat_3, \\
\mu\vvbar_2 &= \onev[]\pibar_2 + \lam_2 \vvbar_{0_1} \Amat_6 + \lam_2 \vvbar_{0_2} \Amat_4, \\
\lam \vvbar_{0_2} &= \onev[]\pibar_{0_2} + \mu \vvbar_1 \Amat_2.
\end{align}
\end{subequations}
We can now use \Thmref{AOI-SHS} to calculate the AoI of source $i$ update in the memory as $\E{x_i} = v_{0_1, i} + v_{0_2, i} + v_{1, i} + v_{2, i}$ for $i \in \{1, 2 \}$ as well as $\E{\hat{x}}= v_{0_1, 3} + v_{0_2, 3} + v_{1, 3} + v_{2, 3}$. 
Some
algebra
yields the following theorem.
\begin{theorem}\label{thm:age-exp}
\mbox{}
\begin{enumerate}
\item[(a)] Source $i$ updates in the memory have average age
\begin{align}
    \E{x_i} = \frac{1}{\mu}\left(\frac{1+\rho}{\rho_i} + \frac{\rho}{1+
    \rho}\right).
    \eqnlabel{source-i-age}
\end{align}
\item[(b)] The max-age process $\hat{x}(t)=\max(x_1(t), x_2(t))$ in the memory has average age
\begin{align}\eqnlabel{Emaxage}
\E{\xhat}&=\frac{(1+\rho)^2(\rho_1^2+\rho_1\rho_2+\rho_2^2)+\rho^2\rho_1\rho_2}{\mu\rho(1+\rho)\rho_1\rho_2}.
\end{align}
\end{enumerate}
\end{theorem}

\begin{figure}[t]
\centering
\includegraphics{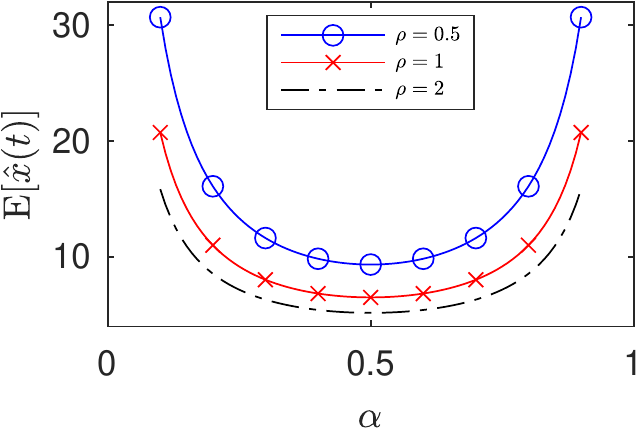}
\caption{Average age of max-age process $\maxage(t)$ in the memory. For a fixed updating load, we vary  $\alpha$ with $\rho_1=\alpha\rho$  and $\rho_2=(1-\alpha)\rho$.
}
\label{fig:maxageplot}
\end{figure}
\noindent
Not surprisingly, the expected max-age $\E{\maxage}$ is symmetric in the  load parameters $\rho_1$ and $\rho_2$. However, since  the formula \eqnref{Emaxage}  is somewhat opaque, a plot of $\E{\maxage}$ appears in Fig.~\ref{fig:maxageplot}. A possibly non-obvious observation from the figure is that increasing the overall updating load $\rho$ generally improves the average max-age because the writer queues no updates. The figure also reveals that the average max-age is penalized by asymmetry in the update rates of the individual sources. This is in part because a source that updates slowly will have high age and thus cause the max-age to be large.  However, it is also true that with asymmetric loads, the high rate  source  will cause updates of the low rate source to be discarded at the writer. Because the writer is non-selective in offering service, it may be performing updates for the high rate source even when the age of that source is already low.

\section{Age of decision updates} \label{sec:decision-processing}
In section~\ref{DPoverview}, we observed that the DP reader is {\em sampling} the source updates from the memory as a point process. 
%
%
%
In particular, we assume the inter-sample times $Y_1,Y_2,\ldots$ that are
i.i.d continuous random variables identical to $Y$. In this case, the update sample times form a renewal process, and in the parlance of \cite{Yates-IT2020}, the update age process $\dpage(t)$ is sampling the max-age update process $\maxage(t)$ in the shared memory.  

\subsection{Average Age at the Decision Process}
At 
time $t$, the most recent read from memory occurred at time $t-Z(t)$. That is,  $Z(t)$ is the age of the sampling renewal process.  When the renewal process is in equilibrium,  $Z(t)$ is stationary with first moment 
\cite[Theorem 5.7.4]{Gallager2013stochastic}
\begin{equation}
\E{Z} = \frac{\E{Y^2}}{2\E{Y}}. \eqnlabel{bcwexc-mom}
\end{equation}
Next, following the approach in \cite{Yates-IT2020}, we observe that the DP reader does not fetch any update in the interval $(t-Z(t),t]$. Hence, at time $t$, the update age $\dpage(t)$ satisfies
\begin{equation}
\dpage(t) = \maxage(t - Z(t)) + Z(t). \eqnlabel{yhat-t}
\end{equation}
Further, $Z(t)$ is independent of $\maxage(t)$ because the inter-sample times $Y_i$ are independent of the age processes in the shared memory. Thus 
 stationarity of $\E{\maxage(t)}$ implies 
 $\E{\maxage(t-Z(t)}=\E{\maxage(t)}=\E{\maxage}$. It then follows from \eqnref{bcwexc-mom} and \eqnref{yhat-t} that $\dpage(t)$ has expected value\footnote{A stronger distributional result is derived in \cite[Theorem~6]{Yates-IT2020} that is not needed for the average AoI analysis here.}  
\begin{equation}
\E{\dpage} = \E{\maxage} + \E{Z}=\E{\maxage} + \frac{\E{Y^2}}{2\E{Y}}. \eqnlabel{Eyhat-Y}
\end{equation}


\subsection{Average Age at the Monitor: Lazy Sampling} 
When the DP reader samples the shared memory, the DP then computes a decision update based on this sample. On delivery of a decision update to the monitor, the update age $\monage(t)$ is reset to the age of the oldest source update that was read and used to compute the decision update. This means that an arrival of decision update at the monitor at time $t$ resets $\monage(t)$ to $\dpage(t)$. 

In this work, we assume that the decision computation times are i.i.d continuous random variables $T_1, T_2, \ldots$, each identically distributed to $T$. We will consider a DP that performs {\em lazy sampling}: after delivering the computation to the output monitor, the DP reader waits for a random time $W$ before reading again. The alternative
    to being lazy is the {\em zero-wait} policy, a special case of lazy when $W=0$.

Fig.~\ref{fig:tc-age-ev} depicts the evolution of the max-age process $\maxage(t)=\max(\oneage(t), \twoage(t))$, the status-sampling process $\dpage(t)$, and the age at the monitor $\monage(t)$.  with i.i.d inter-sample intervals $Y_1, Y_2, \ldots$ such that samples are taken at times $\tau_i = \sum_{j=1}^i Y_j$. 

Under lazy sampling, we admit the possibility that the $i$th computation time $T_i$ and the $i$th waiting time $W_i$ are correlated. However, in order for the $\dpage(t)$ process to be sampling the shared memory with independent inter-sample times $Y_i=T_i+W_i$, we require that the pairs $(T_1,W_1),(T_2,W_2),\ldots$ to be i.i.d., identical to $(T,W)$. Under this assumption, it follows directly from \eqnref{Eyhat-Y} that the average update age at the input to the DP is
\begin{equation}\eqnlabel{DPageTW}
\E{\dpage} = \E{\maxage}+\frac{\E{(T+W)^2}}{2\E{T+W}}.
\end{equation}

Curiously, \eqnref{DPageTW} reveals that the problem of minimizing the average age at the input to DP appears to be isomorphic to the timely updating problem that was originally formulated in \cite{Sun-UBYKS-IT2017UpdateorWait,Yates-isit2015}, where the suboptimality of zero-wait policies was first identified. However, in this system, our objective is not to minimize $\E{\dpage}$ but rather to minimize the average age $\E{\monage}$ at the monitor.  Since $\monage(t)$ is penalized by the waiting time $W$, choosing $W$ to minimize $\E{\dpage}$ may not be good for $\E{\monage}$.
Fortunately, the following claim verifies this is not the case. 



\begin{theorem} \thmlabel{avg-monage-wait}
    If $\maxage(t)$ is a stationary process, 
    then for any  waiting policy such that $W_i$ depends only on $T_i$, the average age 
    at the monitor satisfies 
    \begin{align}
        \E{\monage} &= \E{\dpage} +\E{T}.  \eqnlabel{avg-monage-wait}
    \end{align}
\end{theorem}
The proof appears in the Appendix. We observe that \Thmref{avg-monage-wait}  can give one the mistaken impression that $\E{\monage}$ is insensitive to the waiting time $W$. In fact, the theorem says that the waiting time $W$ affects $\E{\dpage}$ and $\E{\monage}$ identically. A hand-waving intuition is that $\monage(t)$ lags $\dpage(t)$ only during the computation time $T$ but, once the computation is complete, $\monage(t)=\dpage(t)$ during any waiting period.

Combining \eqnref{DPageTW} and \eqnref{avg-monage-wait}, we obtain an end-to-end characterization of the average age in the system:
\begin{align}
        \E{\monage} &= \E{\maxage}+\frac{\E{(T+W)^2}}{2\E{T+W}} +\E{T}.  \eqnlabel{avg-monage-wait-final}
    \end{align}
Since the computation time $T$ is given, \eqnref{avg-monage-wait-final} shows that the choice of a waiting function $W$ as a function of $T$ is the same problem formulated in \cite{Sun-UBYKS-IT2017UpdateorWait,Yates-isit2015}. Hence the solution is the same, namely the $\beta$-minimum waiting policy 
\begin{equation}
    W_i = (\beta - T_i)^+,
\end{equation}
where the parameter $\beta$ is chosen by numerical line search. 
With this policy, $T+W=\max(\beta,T)$ and it follows  from \eqnref{avg-monage-wait-final} that the policy achieves end-to-end  average AoI
\begin{equation}
        \E{\monage} = \E{\maxage} + \frac{\E{\max(\beta^2, T^2)}}{2\E{\max(\beta,T)}} + \E{T}. \eqnlabel{beta-min-age}
    \end{equation}
For completeness, the effectiveness of waiting is demonstrated in  section~\ref{num-eval} by some numerical evaluations of the lazy sampling policy. We will see that lazy sampling becomes important when the variance of the computation time $T$ becomes large.  Before presenting these results, we comment on the connection of this lazy sampling model to the lazy updating model in  \cite{Sun-UBYKS-IT2017UpdateorWait,Yates-isit2015}.

In \cite{Sun-UBYKS-IT2017UpdateorWait,Yates-isit2015}, the random variable
$T$ represented the delivery time of a fresh update (say through a network)  to the monitor. Fresh updates were generated at will and $W$ represented the waiting time prior to generating the next fresh update. A key element of this system was the tight coupling of waiting  and update generation.  In this setting, the intuition behind $\beta$-minimum waiting was that if the prior delivery time was small, the age at the monitor would be small and it would be  a waste of network resources to deliver an update when the age reduction afforded by the update would be small. 

In this work, updates are generated by an exogenous process that is beyond the control of the DP. Moreover, because updates are disseminated through a shared memory publication process, the  age processes of updates in shared memory are essentially uncoupled from the update sampling/processing policy implemented by the DP. In particular, any time the DP reader fetches a sample pair from the memory, the update age of that pair has expected value $\E{\maxage}$, which is just the average age in the shared memory. Nevertheless, even though  DP operation is uncoupled from the age process in shared memory, the $\beta$-minimum waiting policy is effective. In particular, it reduces the expected value of $\dpage(t)$, the age process at the input to the DP. What is happening is that the waiting policy mitigates the deleterious effect of high-variance computation times $T$ on the sampling policy at the DP reader.
We note that \Thmref{avg-monage-wait} went unrecognized in \cite{Sun-UBYKS-IT2017UpdateorWait,Yates-isit2015}. Specifically, \Thmref{avg-monage-wait} shows that no matter what policy is used, the output always lags the input by $\E{T}$ in terms of average age.

\section{Numerical Evaluation}\label{num-eval}
\begin{figure}[t]
\centering
\begin{tabular}{c}
\includegraphics{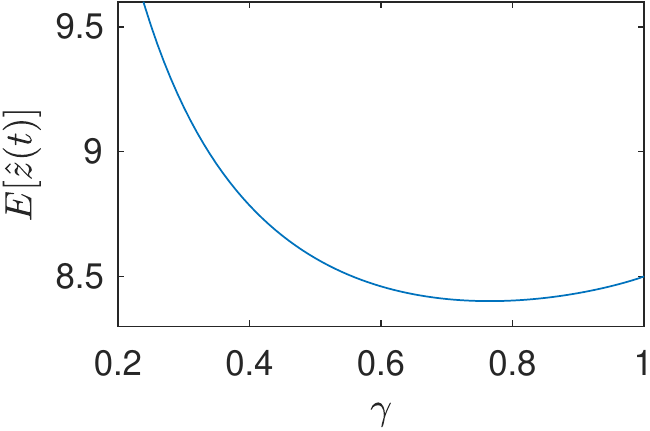} \\
\qquad \textbf{(a)} $T\sim\exp(1)$ \\[3mm]
\includegraphics{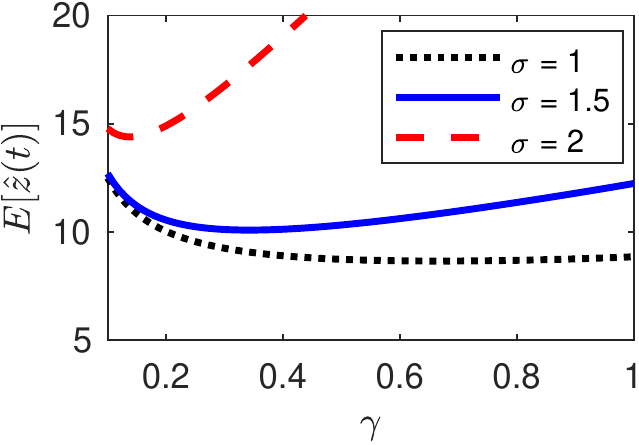} \\
\qquad \textbf{(b)} $T\sim$ Log-Normal($1, e^{\sig^2} - 1$)
\end{tabular}

\medskip

\caption{Average age at the monitor vs the sampling rate $\gamma$ for the $\beta$-minimum policy for different distribution of computation time $T$. Total offered load by source updates is $\rho=1$,
with $\rho_1 = \rho_2 = 0.5$. 
Notice that $\gamma = 1$ is the zero-wait computation policy.}
\label{fig:beta-min-plot}
\end{figure}



In this section, we  examine some numerical examples of the performance $\beta$-minimum waiting policy, simply to remind the reader of the benefits of waiting. 
Fig.~\ref{fig:beta-min-plot} illustrates age performance with respect to variance in the computation time with probability distributions $\exp(1)$ (Fig.~\ref{fig:beta-min-plot}(a)), and Log-Normal($1, e^{\sig^2} - 1$) (Fig.~\ref{fig:beta-min-plot}(b)). 
The log-normal distributed computation times $T$ has PDF \cite{yates2014probability},
\begin{align}
f_T(t) &= 
\frac{e^{-(\ln(t) - b)^2/2\sig^2}}{\sqrt{2\pi}\sig t}, \qquad t > 0,
\end{align}
with free parameters $b$ and $\sig > 0$.
In our numerical evaluations, 
we consider a given distribution on $T$ such that the computation time is normalized to $\E{T}=1$. In this regard, for Log-Normal distribution,
for each $\sig$, we set $b = -\sig^2/2$ so that $\E{T} = 1$. By varying $\sigma$, we vary $\Var{T}=e^{\sigma^2}-1$.
These numerical results are largely similar to those in \cite{Sun-UBYKS-IT2017UpdateorWait,Yates-isit2015}.
In particular, the results remind the reader that zero-wait becomes increasingly sub-optimal when the computation time $T$ has high variance. 
The choice of $\beta$ specifies a sampling rate  
\begin{equation}
\gamma = \frac{1}{\E{T} + \E{W}} =\frac{1}{\E{\max(\beta,T)}}\eqnlabel{lam}
\end{equation}
at the DP reader. We then plot the average age at the monitor as a function of $\gamma$.  Because $\E{T}=1$, the maximum update sampling rate is $\gamma=1$, which corresponds to the zero-wait policy. As $\gamma\to 0$, the average age is increasingly dominated by the average inter-read time  $1/\gamma$, because  updates become too infrequent. 



\section{Conclusion}
\label{sec:conclusion}
In this work, we focused on the problem of
timely processing of updates from multiple sources.  
Specifically, we considered a model of a publish-subscribe system where a writer publishes updates from two independent sources in a shared memory and decision updates are derived by a decision process by reading from the memory. 
The decision processing is a subscriber that works independently of how the source updates are recorded in the memory. 
Even though the decision processing operates without knowledge of the ages of updates in the shared memory,
its reading policy is  still able to improve the end-to-end decision update timeliness.

We recognize that there could be other DP reading policies that exploit knowledge of the update age processes in shared memory to further reduce decision update age at the monitor;  identifying such policies 
would be an interesting avenue for future research.

Another drawback of our system model is  that when  stale values are read from the shared memory, the DP will still perform its computations, even though the resulting DP update is not age-reducing. Effectively, the DP is waiting for a computation time before attempting to retrieve new updates from memory. Instead, it would be better for the DP to discard the stale updates and wait for an optimized time before reading the memory again. Analyzing such a model also remains as future work. 

\section*{Acknowledgment}
This work was supported by the US National Science Foundation under grant number CNS-2148104.

\bibliographystyle{IEEEtran}
\bibliography{refs,AOI-2020-03}

\appendix

\begin{proof} (\Thmref{avg-monage-wait})
     Suppose $t\in (\tau_{i-1},\tau_i]$, the $i$th inter-sample interval. 
    We observe from Fig.~\ref{fig:tc-age-ev} that $\dpage(\tau_{i-1})=\maxage(\tau_{i-1})$ because the reader fetches update $i-1$ at that time.  However, at that time, the monitor has only received the decision update based on 
    update $i-2$, which had age $\dpage(\tau_{i-2})$ at time $\tau_{i-2}$ and now, at time $\tau_{i-1}=\tau_{i-2}+Y_{i-1}$, has age $\dpage(\tau_{i-2})+Y_{i-1}$.  Hence, at time $\tau_{i-1}$, the monitor has age
     \begin{equation}
       \monage(\tau_{i-1})=\dpage(\tau_{i-2})+Y_{i-1}.
       \end{equation}
       Defining $H_i=\dpage(\tau_{i-2}) + Y_{i-1} - \dpage(\tau_{i-1})$, we can write       
\begin{equation}
\monage(\tau_{i-1})   =\dpage(\tau_{i-1})+ H_i.
\end{equation}
Since $\dpage(\tau_i)=\maxage(\tau_i)$ for all $i$, \begin{equation}
 H_i=\maxage(\tau_{i-2}) + Y_{i-1} - \maxage(\tau_{i-1}).
\end{equation}
It follows from  stationarity of $\maxage(t)$ and independence of the sampling times $\tau_i$  and $\maxage(t)$ that
\begin{align}
    \E{H_i}&=\E{\maxage(\tau_{i-2})} + \E{Y_{i-1}} - \E{\maxage(\tau_{i-1})}\nn
    &= \E{Y_{i-1}}=\E{T}+\E{W}.\eqnlabel{ehi} 
\end{align}
At time $\tau_{i-1}$, the $i$th busy period starts and both $\dpage(t)$ and $\monage(t)$ grow linearly at rate $1$ because neither process sees an update. Hence, $\monage(t)   =\dpage(t)+ H_i$ during the busy period. Only when the busy period completes at time $\tau_{i-1}+T_i$ does $\monage(t)$ drop and become equal to $\dpage(t)$. 
Let events $B_t$ and $I_t$ correspond to the decision process being busy and idle respectively, at time $t$.
In this interval, the event $B_t$ occurs while $\tau_{i-1} \leq t \leq \tau_{i-1} + T_i$; otherwise $I_t$ occurs if $\tau_{i-1} + T_i \leq t \leq \tau_{i}$. With these events, we can write
\begin{align}
        \monage(t) &= \begin{cases}
            \dpage(t) + H_i, & \text{if $B_t$}, \\
            \dpage(t), & \text{if $I_t$}.
        \end{cases}
    \end{align}

For $t\ge \tau_{i-1}$, event $B_t$ is independent of $H_i$, and it follows from  the law of total expectation that
  \begin{align}
        \E{\monage(t)} &= \E{\dpage(t) +H _i| B}\prob{B_t} + \E{\dpage(t) | I_t} \prob{I_t} \nonumber \\
        &= \E{\dpage(t)} + \E{H_i}\prob{B_t}.
        \eqnlabel{ezhat}
    \end{align}
    In each renewal period, the decision process is busy for time $T$ and then idle for time $W$. By considering a renewal reward process in which a reward  $T$ is earned for the busy period, it follows that the limiting fraction of time spent in a busy state is given by
    \begin{equation}
        \prob{B_t} = \frac{\E{T}}{\E{T} + \E{W}}. \eqnlabel{pb}
    \end{equation}
    Applying \eqnref{ehi} and \eqnref{pb} to \eqnref{ezhat} yields the claim.
\end{proof}

\end{document}